\documentclass[11pt,a4paper]{article}

\usepackage{moriond}
\usepackage{graphicx}
\usepackage{amsmath}
\usepackage{amssymb}
\usepackage{amsfonts}
\usepackage{amscd}
\usepackage{mathrsfs}
\usepackage{bbm}

\def\Journal#1#2#3#4{{#1} {\bf #2}, #3 (#4)}

\def\NPB{{\em Nucl. Phys.} B}
\def\PLB{{\em Phys. Lett.} B}
\def\PRD{{\em Phys. Rev.} D}
\def\JHEP{{\em JHEP}}

\newcommand{\D}{\ensuremath{\mathrm{d}}}

\newcommand{\eVdist}{\kern-0.06667em}
\newcommand{\gev}{{\,\text{Ge}\eVdist\text{V\/}}}
\newcommand{\tev}{{\,\text{Te}\eVdist\text{V\/}}}

\newcommand{\ps}{\text{\sc ps}}
\newcommand{\GG}{\text{\sc gg}}
\newcommand{\sm}{\text{\sc sm}}

\allowdisplaybreaks[1]

\begin{document}

\begin{flushright}
DESY 06-067
\end{flushright}

\vspace*{4cm}
\title{ 
Squarks and Sleptons between Branes and Bulk --\\
Gaugino Mediation and Gravitino Dark Matter in an SO(10) Orbifold GUT
\footnote{Talk presented at the XLIst Rencontres de Moriond,
 March 11-18, 2006, La Thuile, Italy.  Based on work done in
 collaboration with Wilfried Buchm\"uller and Kai Schmidt-Hoberg
 \cite{Buchmuller:2005ma}.
}
}

\author{J.~KERSTEN}

\address{DESY Theory Group, Notkestr.~85,\\
22603 Hamburg, Germany}

\maketitle\abstracts{
We study gaugino-mediated supersymmetry breaking in a six-dimensional $SO(10)$
orbifold GUT model where quarks and leptons are mixtures of brane and
bulk fields.  The couplings of bulk matter fields to the supersymmetry
breaking brane field have to be suppressed in order to avoid large
FCNCs.  We derive bounds on the soft supersymmetry breaking parameters
and calculate the superparticle mass spectrum.  If the gravitino is the
LSP, the $\tilde\tau_1$ or the $\tilde\nu_{\tau\mathrm{L}}$ turns out to
be the NLSP, with characteristic signatures at future colliders and in
cosmology.
}

\section{Introduction}
Supersymmetric orbifold GUTs are attractive candidates for
unified theories explaining the masses and mixings of fermions, see for
example \cite{Kawamura:1999nj,Hall:2001pg,Hebecker:2001wq}.
Features like doublet-triplet splitting and absence of
dimension-five proton decay operators, which are difficult to
realise in four-dimensional grand unified theories, are easily obtained.
In order to add predictions for the superparticle mass spectrum, an
orbifold model has to be supplemented by a scenario for SUSY breaking.
Given the higher-dimensional setup with various branes, this scenario
involves in general both bulk and brane fields.

According to this reasoning, we combine an $SO(10)$ theory in six
dimensions, proposed in~\cite{Asaka:2003iy}, with
gaugino-mediated SUSY breaking \cite{Kaplan:1999ac,Chacko:1999mi}.  The
orbifold compactification of the two extra dimensions has four fixed
points or ``branes''.  On three of them, three quark-lepton generations
are localised.  The Standard Model leptons and down-type quarks are
linear combinations of these localised fermions
and a partial fourth generation living in the bulk.  This leads to the
observed large neutrino mixings.  On the fourth brane, a
field $S$ develops an $F$-term vacuum expectation value
(vev) breaking SUSY.  As the gauge and Higgs fields propagate in
the bulk, they feel the effects of SUSY breaking.  Thus, gauginos and
Higgs scalars obtain soft masses.  The soft masses and trilinear
couplings of
the scalar quarks and leptons approximately vanish at the
compactification scale.  Non-zero values are generated by the running to
low energies, which leads to a realistic superparticle mass spectrum.
If the gravitino is the lightest superparticle (LSP), it can form
the dark matter.  The next-to-lightest
superparticle (NLSP) is then a scalar tau or a scalar neutrino, which is
consistent with constraints from big bang nucleosynthesis.

\section{The Orbifold GUT Model}
We consider an $N=1$
supersymmetric $SO(10)$ gauge theory in 6 dimensions 
compactified on the orbifold $T^2/\left(
      {\mathbbm Z}_2 \times {\mathbbm Z}_2^\prime  \times 
{\mathbbm Z}_2^{\prime\prime} \right)$ \cite{Asaka:2003iy}.
The theory has 4 fixed points,
$O_\text{\sc i}$, $O_{\ps}$, $O_{\GG}$ and $O_\text{fl}$, located at
the corners of a ``pillow'' corresponding to the two compact
dimensions.  At $O_\text{\sc i}$ the full $SO(10)$ survives, 
whereas at the other fixed points $SO(10)$ is broken to its
subgroups \mbox{${ G}_{\ps}={ SU(4)}\times { SU(2)} \times
  { SU(2)}$}, ${ G}_{\GG}={ SU(5)}\times {
  U(1)}_X$ and flipped $SU(5)$,
\mbox{${ G}_\text{fl}={ SU(5)'}\times { U(1)'}$},
respectively. The intersection of these GUT groups yields the
Standard Model group with an additional $U(1)$ factor, 
$G_{\sm '}= SU(3) \times SU(2) \times U(1)_Y \times U(1)_X$, 
as unbroken gauge symmetry below the
compactification scale, which we identify with the GUT scale.

The field content of the theory is strongly constrained by requiring
the cancellation of bulk and brane anomalies.
The brane fields are the {\bf 16}-plets
$\psi_1,\psi_2,\psi_3$.  The bulk contains
six {\bf 10}-plets, $H_1,\dots, H_6$, and four {\bf 16}-plets, 
$\Phi, \Phi^c, \phi, \phi^c$, as hypermultiplets.  Vevs
of $\Phi$ and $\Phi^c$ break the surviving 
${U(1)}_{B-L}$.  The electroweak gauge group is broken by expectation
values of the doublets contained in $H_1$ and $H_2$.
The zero modes of $\phi,\phi^c$ and $H_5,H_6$,
act as a fourth generation of down quarks
and leptons and mix with the three generations of brane fields.  We
allocate these {\bf 16}-plets to the branes where
$SO(10)$ is broken, placing $\psi_1$ at
$O_{\GG}$, $\psi_2$ at $O_\text{fl}$ and $\psi_3$ at $O_{\ps}$.  The three
``families'' are then separated by distances large compared to the
cutoff scale $\Lambda$. Hence, they can only have diagonal Yukawa
couplings with the bulk Higgs fields. The brane fields, however, can mix
with the bulk zero modes without suppression. As these mixings take place
only among left-handed leptons and right-handed down-quarks, we
obtain a characteristic pattern of mass matrices.
The allowed terms in the superpotential are restricted by $R$-invariance
and an additional $U(1)_{\tilde{X}}$ symmetry with the charge
assignments given in Tab.~\ref{tab:charge}.
The most general superpotential satisfying these constraints
is given in \cite{Asaka:2003iy}.
It determines the SUSY-conserving mass terms and Yukawa couplings.

\begin{table}
  \caption{Charge assignments for the symmetries $U(1)_R$ 
           and $U(1)_{\tilde{X}}$
    \label{tab:charge}}
  \centering
  \renewcommand{\arraystretch}{1.1}
  \begin{tabular}{|c||c|c|c|c|c|c|c|c|c|c|c|c|c|}
    \hline
     & $H_1$ & $H_2$ & $\Phi^c$ & $H_3$ & $\Phi$ & 
       $H_4$ & $\psi_i$ & $\phi^c$ & $\phi$ & $H_5$ & $H_6$ & $S$ \\
    \hline \hline
$R$ & 0 & 0 & 0 & 2 & 0 & 2 & 1 & 1 & 1 & 1 & 1 & 0 \\
    \hline 
$\tilde{X}$ & -2a & -2a & -a & 2a & a & -2a & a & -a & a & 2a & -2a & 0\\
    \hline
  \end{tabular}
\end{table}

Soft SUSY-breaking terms are generated by gaugino mediation
\cite{Kaplan:1999ac,Chacko:1999mi}.  A gauge-singlet chiral superfield
$S$, which is localised at the fixed point $O_\text{\sc i}$, acquires a
non-vanishing vev for its $F$-term component.  Supersymmetry is then
fully broken and the breaking can be communicated to bulk fields by
direct interactions.  In the case of the gauginos, these are of the form
\begin{equation}
	\mathscr{L}_S \supset 
	\frac{g_4^2 h}{4 \Lambda}\,\int \D^2\theta \,S \,W^\alpha W_\alpha 
	+ \text{h.c.} \;,
\end{equation}
where $g_4$ is the four-dimensional gauge coupling and $h$ is a
dimensionless coupling.
Further interactions that are relevant for SUSY breaking and respect all
symmetries are obtained by multiplying terms in the superpotential by
$S/\Lambda$.

Soft masses for the Higgses and for all bulk matter fields, as well as a
$\mu$- and a $B\mu$-term arise from the K\"ahler potential.  In order to
obtain a $\mu$-term, we assume the global $U(1)_{\tilde{X}}$ symmetry to
be only approximate and allow for explicit breaking here.  Although the
$\mu$-term itself is not a soft term, it is thus generated only after
SUSY breaking via the Giudice-Masiero mechanism
\cite{Giudice:1988yz}.

The MSSM squarks and sleptons live on different branes than $S$.
Therefore, they obtain soft masses only via loop contributions
through the bulk, which are negligible here, and via renormalisation
group running.

\section{The Scalar Mass Matrices and FCNCs}
From the superpotential one obtains
$4\times 4$ matrices of the form
\begin{equation} \label{eq:matrix-structure}
	m =
	\left(\begin{array}{cccc}
	 \mu_1 & 0 & 0 & \widetilde{\mu}_1 \\
	 0 & \mu_2 & 0 & \widetilde{\mu}_2 \\
	 0 & 0 & \mu_3 & \widetilde{\mu}_3 \\
	 \widetilde{M}_1 & \widetilde{M}_2 & \widetilde{M}_3 & \widetilde{M}_4
	\end{array} \right)
\end{equation}
for the down-type quark, charged lepton and Dirac neutrino masses.
Here $\mu_i, \widetilde{\mu}_i \sim v$ and 
$\widetilde{M}_i \sim M_\mathrm{GUT}$.  While $\mu_i$ and
$\widetilde{\mu}_i$ have to be hierarchical, we assume no hierarchy
between the $\widetilde{M}_i$.
The up-type quark and Majorana neutrino mass matrices are diagonal
$3\times3$ matrices, since the corresponding fields do not have partners
in the bulk.
At the compactification scale, we integrate out the heavy degrees of
freedom to obtain an effective theory with three generations.  This
requires block-diagonalising the mass matrices $m$ by transformations
involving unitary matrices $U_4$ and $V_4$.  In the case of the leptons,
$V_4$ contributes to the observed large mixing between the left-handed
fields.  On the other hand, $U_4$ is close to the unit matrix, so that
there is only small mixing among the right-handed fields.  The situation
is reversed in the down-quark sector, where the right-handed fields are
strongly mixed while the left-handed ones are not.

Only the scalars of the fourth
generation, which are very heavy, obtain soft masses, since they are
bulk fields.  However, the transformations diagonalising the fermion
mass matrices transmit SUSY-breaking effects from the fourth to the
light generations.  As some of them cause large mixing, there are soft
mass matrices whose off-diagonal elements are generically of similar
size as the diagonal elements in a basis where quark and lepton mass
matrices are diagonal.  This leads to unacceptably large
flavour-changing neutral currents.
We expect this problem to be generic in higher-dimensional
theories with mixing between bulk and brane matter fields as long as the
bulk fields can couple to the hidden sector.
In the following, we shall
assume that soft masses for bulk matter fields, contrary to the bulk
Higgs fields, are strongly suppressed.

\section{The Low-Energy Sparticle Spectrum}
Imposing vanishing soft masses for the bulk matter fields,
the boundary conditions
at the compactification scale are those of the usual gaugino mediation
scenario with bulk Higgs fields \cite{Chacko:1999mi},
\begin{align}
	g_1 &= g_2 = g_3 = g \simeq \frac{1}{\sqrt{2}} \;,
\\
	M_1 &= M_2 = M_3 = m_{1/2} \;,
\\
	m_{\tilde \phi_\mathrm{L}}^2 =
	m_{\tilde \phi_\mathrm{R}}^2 &= 0
	\quad \text{for all squarks and sleptons } \tilde\phi \;,
\\
	A_{\tilde\phi} &= 0
	\quad \text{for all squarks and sleptons } \tilde\phi\;,
\\
	\mu, B\mu, m^2_{\tilde{h}_i} &\neq 0 \quad (i=1,2) \;.
\end{align}

As a benchmark point for our discussion, we choose $m_{1/2} = 500\gev$,
$\tan\beta=10$ and $\text{sign}(\mu)=+1$.  The
LSP can be the gravitino, with a mass between 50 and $100\gev$
\cite{Buchmuller:2005rt}.
The values of $\mu$ and $B\mu$ are determined by the
conditions for electroweak symmetry breaking.
In order to find the spectrum at low energy, we have to take into
account the running of the parameters.  We employ SOFTSUSY
\cite{Allanach:2001kg} for this purpose.

The 1-loop running of the gaugino masses does not depend on
the scalar masses, so that their low-energy values remain virtually
the same in all cases as long as we do not change $m_{1/2}$.
Numerically, we find
$M_1(M_Z) \simeq 200\gev$,
$M_2(M_Z) \simeq 380\gev$, and
$M_3(M_Z) \simeq 1200\gev$.
To good approximation, the lightest neutralino is the bino and the
second-lightest one is the wino, unless the soft Higgs mass
$m^2_{\tilde{h}_2}$ is sizable.  In
the latter case, the electroweak symmetry breaking conditions lead to a
rather small $\mu$, so that there is
significant mixing between the neutralinos.

\begin{figure}[p]
\centering
\includegraphics{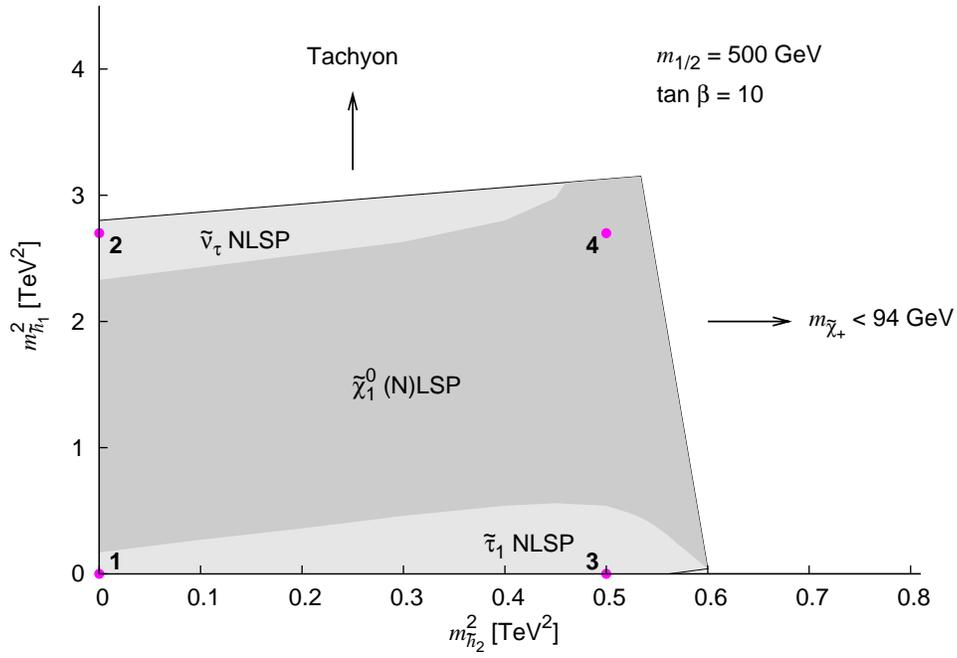}
\vspace{-3mm}
\caption{Allowed region for the soft Higgs masses.  In the dark-gray
 area, a neutralino is lighter than all sleptons.  For the points marked
 by the dots, the resulting superparticle mass spectrum is
 shown in Fig.~\ref{fig:LineSpectra}.} 
\label{fig:parameterspace}
\end{figure}

One constraint on the model parameters is that the running
down to the weak scale must not produce tachyons.
This yields an upper bound on $m_{\tilde{h}_1}^2$.
The bound on $m_{\tilde{h}_2}^2$ is due to the experimental
limits on the superparticle masses \cite{Eidelman:2004wy}.  If the
initial value of $m_{\tilde{h}_2}^2$ is too large, this mass squared 
crosses zero at a rather low energy, so that its absolute value at the
electroweak scale is small.  Consequently, the $\mu$ parameter is also
small, leading to a Higgsino-like chargino with a mass below the current
limit of $94\gev$.  This limit on
$m_{\tilde{h}_2}^2$ is the relevant one for almost all values of  
$m_{\tilde{h}_1}^2$.  Only for very small $m_{\tilde{h}_1}^2$, the 
experimental requirement that the $\tilde\tau_1$ be heavier than $86\gev$
becomes more restrictive.
The resulting allowed region in parameter space is the
gray-shaded area in Fig.~\ref{fig:parameterspace}.

\begin{figure}[p]
\centering
\vspace{1mm}
\includegraphics[width=0.98\linewidth]{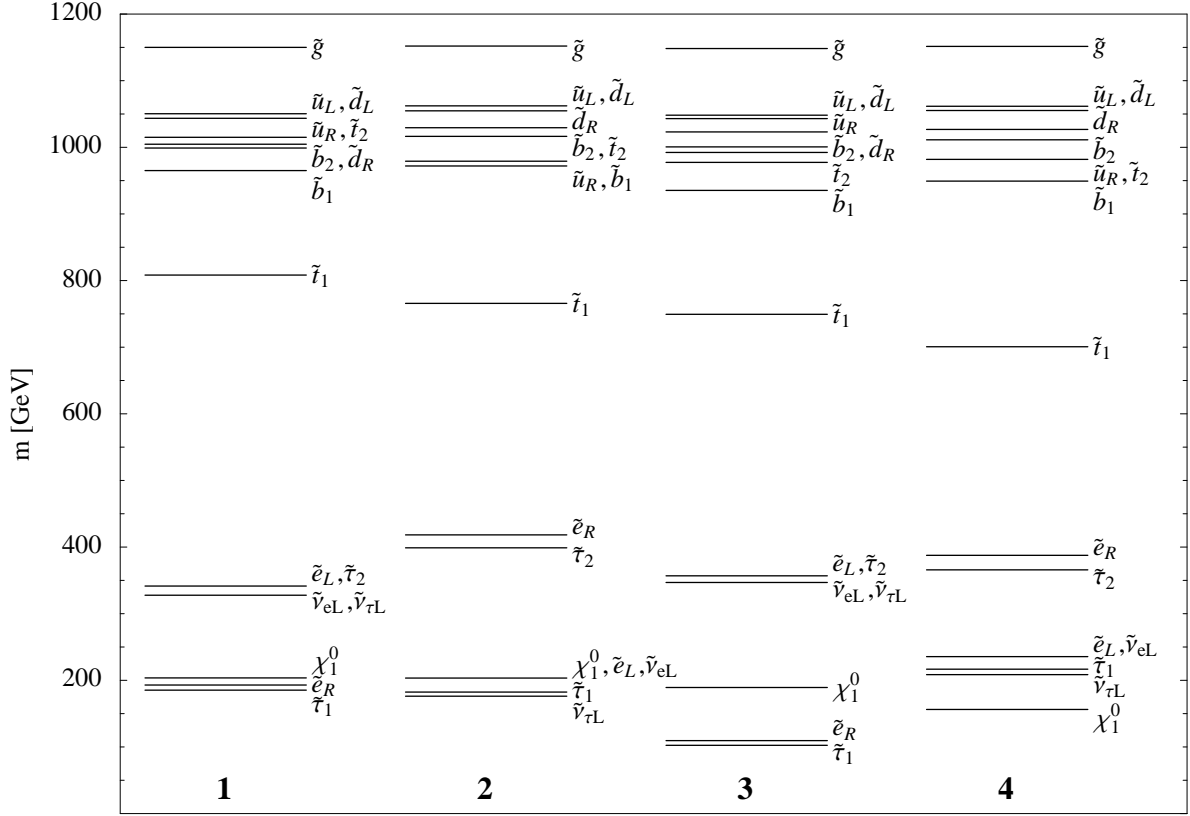}
\caption{
 Spectra of superparticle pole masses.  The numbers at the bottom
 correspond to the points in parameter space marked by the coloured dots
 in Fig.~\ref{fig:parameterspace}.  The high-energy boundary conditions
 for the soft Higgs masses were
 $m^2_{\tilde{h}_1}=m^2_{\tilde{h}_2}=0$ (point 1),
 $m^2_{\tilde{h}_1}=2.7\tev^2$, $m^2_{\tilde{h}_2}=0$ (point 2),
 $m^2_{\tilde{h}_1}=0$, $m^2_{\tilde{h}_2}=0.5\tev^2$ (point 3), and
 $m^2_{\tilde{h}_1}=2.7\tev^2$, $m^2_{\tilde{h}_2}=0.5\tev^2$ (point 4),
 respectively.  In all cases, we used $m_{1/2}=500\gev$, $\tan\beta=10$
 and $\text{sign}(\mu)=+1$.  As the first and second generation scalars
 are degenerate, only the first generation is listed in the figure.
 Particles with a mass difference of less than about $3\gev$ are
 represented by a single line.  The heavier neutralinos and the
 charginos have been omitted for better readability.
}
\label{fig:LineSpectra}
\end{figure}

In Fig.~\ref{fig:LineSpectra}, we show the superparticle spectra we
obtain at the 4 points in parameter space marked by dots
in Fig.~\ref{fig:parameterspace}.
Due to the large effects of the strong interaction, the squark masses
experience the fastest running and end up around a TeV.  
If all scalar soft masses vanish at the GUT scale (point 1), the
left-handed slepton masses change significantly in the beginning, but
afterwards the evolution flattens.  They reach
values between $300$ and $400\gev$ at low energies.  The flattening of the evolution is
even more pronounced for the right-handed slepton masses.
As a consequence, these scalars remain lighter than the lightest
neutralino \cite{Kaplan:1999ac}.
This is also the case for $m^2_{\tilde{h}_2} > m^2_{\tilde{h}_1}$ (point
3), since then the evolution of the
right-handed slepton masses is slowed down further, while that of the
left-handed masses is enhanced.

For $m^2_{\tilde{h}_1} > m^2_{\tilde{h}_2}$, important changes can
occur \cite{Chacko:1999mi,Kaplan:2000av,Schmaltz:2000ei},
in particular in the slepton
spectrum.  For the largest possible difference of the soft Higgs masses,
the left-handed sleptons remain relatively light, with a low-energy mass
below $200\gev$.  Contrary to that, the right-handed slepton masses run
unusually fast near the GUT scale and reach values close to $400\gev$ at
low energy.  Thus, the NLSP is a sneutrino in this case
\cite{Kaplan:2000av}, with a slightly heavier stau $\tilde\tau_1$ (cf.\
point~2).
If $m^2_{\tilde{h}_1}$ is neither close to zero nor to its upper bound
(point 4),
the running of the right-handed slepton masses is sufficiently enhanced
to lift them above the lightest neutralino mass.  At the same time, the
running of the left-handed slepton masses is damped weakly enough, so
that they are heavier than the lightest neutralino, too
\cite{Chacko:1999mi,Kaplan:2000av}.  A neutralino NLSP together with a
gravitino LSP heavier than a GeV is excluded by cosmology, see e.g.\
\cite{Cerdeno:2005eu} for the most recent analysis.
Therefore, this case is only viable if the neutralino is the LSP while the
gravitino is heavier.  This is possible, because we only have a lower
bound on the gravitino mass in gaugino mediation.  The corresponding region in parameter
space is the dark-gray area in Fig.~\ref{fig:parameterspace}.

Varying the high-energy gaugino mass simply leads to a rescaling of the
scalar spectrum to a first approximation.  If $m_{1/2}$ is increased
while keeping the other soft masses fixed, the spectrum comes closer to
the one obtained in the minimal case of vanishing scalar masses.  The
LEP bound on the lightest Higgs mass leads to a lower bound on
$m_{1/2}$.  If $m^2_{\tilde{h}_1}$ takes its maximal value, a unified
gaugino mass of slightly less than $400\gev$ is compatible with the LEP
bound (for $m_t=172.7\gev$).

A change of $\tan\beta$ leads to a change of the mass splitting between
the third generation and the first two.
If $\tan\beta$ is significantly smaller than 10, the value used in our
benchmark scenario, the Higgs mass bound leads to severer restrictions.
If $\tan\beta<6$, this bound is violated even for maximal $m_t$ and
$m^2_{\tilde{h}_1}$, i.e.\ a gaugino mass larger than $500\gev$ is
required.
For larger values of $\tan\beta$, the lighter stau mass decreases a lot
faster at lower energies.  Hence, the parameter space region shrinks
where the lightest neutralino is lighter than the $\tilde\tau_1$.  For
$\tan\beta=25$, this region almost vanishes.  On the other hand, the
soft Higgs masses have to satisfy severer upper bounds in order to
avoid tachyons and a too light stau.  For $\tan\beta=35$, the model is
only viable if all soft scalar masses vanish at the GUT scale, and for
$\tan\beta>35$ the lighter stau mass always lies below its experimental
limit unless the gauginos are heavier than $500\gev$.
We conclude that the model favours $10 \lesssim \tan\beta \lesssim 25$.

\section{Conclusions}
We have discussed gaugino-mediated SUSY breaking in a six-dimensional
$SO(10)$ orbifold GUT where quarks and leptons are mixtures of brane
and bulk fields.  The couplings of bulk matter fields to the SUSY
breaking brane field have to be suppressed in order to
avoid flavour-changing neutral currents.  The compatibility of the SUSY breaking mechanism and
orbifold GUTs with brane and bulk matter fields is a generic problem
which requires further studies.

The parameters relevant for the superparticle mass spectrum are the
universal gaugino mass, the
soft Higgs masses, $\tan\beta$ and the sign of $\mu$.  We have analysed
their impact on the spectrum and determined the region in parameter
space that results in a viable phenomenology.  The model favours
moderate values of $\tan\beta$ between about 10 and 25.  The gaugino
mass at the GUT scale should not be far below $500\gev$ in order to
satisfy the LEP bound on the Higgs mass.  Typically, the lightest
neutralino is bino-like with a mass of $200\gev$, and the gluino mass is
about $1.2\tev$.  Either the right-handed or the left-handed sleptons
can be lighter than the neutralinos.  The corresponding region in
parameter space grows with $\tan\beta$.  In this region, the gravitino
is the LSP with a mass around $50\gev$.  The $\tilde\tau_1$ or the
$\tilde\nu_{\tau\mathrm{L}}$ is the NLSP.
A sneutrino NLSP has the advantage that constraints from big bang
nucleosynthesis and the cosmic microwave background are less stringent
\cite{Feng:2004mt}.  For a stau NLSP, on the other hand,
there exists
the exciting possibility that its decays may lead to the discovery of
the gravitino in future collider experiments
\cite{Buchmuller:2004rq}.

\section*{References}
\frenchspacing

\providecommand{\bysame}{\leavevmode\hbox to3em{\hrulefill}\thinspace}

\end{document}